\documentclass[longauth, traditabstract, letter]{aa}  

\usepackage{graphicx}
\usepackage{natbib}
\usepackage{txfonts}
\begin{document}
   \title{HerMES: SPIRE galaxy number counts at 250,
350 and 500\,$\mu$m\thanks{{\it Herschel} is an ESA space observatory with science instruments provided
by European-led Principal Investigator consortia and with important participation from NASA.}}

\author{S.\,J.~Oliver\inst{1}
\and L.~Wang\inst{1}
\and A.\,J.~Smith\inst{1}
\and B.~Altieri\inst{2}
\and A.~Amblard\inst{3}
\and V.~Arumugam\inst{4}
\and R.~Auld\inst{5}
\and H.~Aussel\inst{6}
\and T.~Babbedge\inst{7}
\and A.~Blain\inst{8}
\and J.~Bock\inst{8,9}
\and A.~Boselli\inst{10}
\and V.~Buat\inst{10}
\and D.~Burgarella\inst{10}
\and N.~Castro-Rodr{\'\i}guez\inst{11,12}
\and A.~Cava\inst{11,12}
\and P.~Chanial\inst{7}
\and D.\,L.~Clements\inst{7}
\and A.~Conley\inst{13}
\and L.~Conversi\inst{2}
\and A.~Cooray\inst{3,8}
\and C.\,D.~Dowell\inst{8,9}
\and E.~Dwek\inst{14}
\and S.~Eales\inst{5}
\and D.~Elbaz\inst{6}
\and M.~Fox\inst{7}
\and A.~Franceschini\inst{15}
\and W.~Gear\inst{5}
\and J.~Glenn\inst{13}
\and M.~Griffin\inst{5}
\and M.~Halpern\inst{16}
\and E.~Hatziminaoglou\inst{17}
\and E.~Ibar\inst{18}
\and K.~Isaak\inst{5}
\and R.\,J.~Ivison\inst{18,4}
\and G.~Lagache\inst{19}
\and L.~Levenson\inst{8,9}
\and N.~Lu\inst{8,20}
\and S.~Madden\inst{6}
\and B.~Maffei\inst{21}
\and G.~Mainetti\inst{15}
\and L.~Marchetti\inst{15}
\and K.~Mitchell-Wynne\inst{3}
\and A.\,M.\,J.~Mortier\inst{7}
\and H.\,T.~Nguyen\inst{9,8}
\and B.~O'Halloran\inst{7}
\and A.~Omont\inst{22}
\and M.\,J.~Page\inst{23}
\and P.~Panuzzo\inst{6}
\and A.~Papageorgiou\inst{5}
\and C.\,P.~Pearson\inst{24,25}
\and I.~P{\'e}rez-Fournon\inst{11,12}
\and M.~Pohlen\inst{5}
\and J.\,I.~Rawlings\inst{23}
\and G.~Raymond\inst{5}
\and D.~Rigopoulou\inst{24,26}
\and D.~Rizzo\inst{7}
\and I.\,G.~Roseboom\inst{1}
\and M.~Rowan-Robinson\inst{7}
\and M.~S\'anchez Portal\inst{2}
\and R.~Savage\inst{1,27}
\and B.~Schulz\inst{8,20}
\and Douglas~Scott\inst{16}
\and N.~Seymour\inst{23}
\and D.\,L.~Shupe\inst{8,20}
\and J.\,A.~Stevens\inst{28}
\and M.~Symeonidis\inst{23}
\and M.~Trichas\inst{7}
\and K.\,E.~Tugwell\inst{23}
\and M.~Vaccari\inst{15}
\and E.~Valiante\inst{16}
\and I.~Valtchanov\inst{2}
\and J.\,D.~Vieira\inst{8}
\and L.~Vigroux\inst{22}
\and R.~Ward\inst{1}
\and G.~Wright\inst{18}
\and C.\,K.~Xu\inst{8,20}
\and M.~Zemcov\inst{8,9}}

\institute{Astronomy Centre, Dept. of Physics \& Astronomy, University of Sussex, Brighton BN1 9QH, UK\\
 \email{S.Oliver@Sussex.ac.uk}
\and Herschel Science Centre, European Space Astronomy Centre, Villanueva de la Ca\~nada, 28691 Madrid, Spain
\and Dept. of Physics \& Astronomy, University of California, Irvine, CA 92697, USA
\and Institute for Astronomy, University of Edinburgh, Royal Observatory, Blackford Hill, Edinburgh EH9 3HJ, UK
\and Cardiff School of Physics and Astronomy, Cardiff University, Queens Buildings, The Parade, Cardiff CF24 3AA, UK
\and Laboratoire AIM-Paris-Saclay, CEA/DSM/Irfu - CNRS - Universit\'e Paris Diderot, CE-Saclay, pt courrier 131, F-91191 Gif-sur-Yvette, France
\and Astrophysics Group, Imperial College London, Blackett Laboratory, Prince Consort Road, London SW7 2AZ, UK
\and California Institute of Technology, 1200 E. California Blvd., Pasadena, CA 91125, USA
\and Jet Propulsion Laboratory, 4800 Oak Grove Drive, Pasadena, CA 91109, USA
\and Laboratoire d'Astrophysique de Marseille, OAMP, Universit\'e Aix-marseille, CNRS, 38 rue Fr\'ed\'eric Joliot-Curie, 13388 Marseille cedex 13, France
\and Instituto de Astrof{\'\i}sica de Canarias (IAC), E-38200 La Laguna, Tenerife, Spain
\and Departamento de Astrof{\'\i}sica, Universidad de La Laguna (ULL), E-38205 La Laguna, Tenerife, Spain
\and Dept. of Astrophysical and Planetary Sciences, CASA 389-UCB, University of Colorado, Boulder, CO 80309, USA
\and Observational  Cosmology Lab, Code 665, NASA Goddard Space Flight  Center, Greenbelt, MD 20771, USA
\and Dipartimento di Astronomia, Universit\`{a} di Padova, vicolo Osservatorio, 3, 35122 Padova, Italy
\and Department of Physics \& Astronomy, University of British Columbia, 6224 Agricultural Road, Vancouver, BC V6T~1Z1, Canada
\and ESO, Karl-Schwarzschild-Str. 2, 85748 Garching bei M\"unchen, Germany
\and UK Astronomy Technology Centre, Royal Observatory, Blackford Hill, Edinburgh EH9 3HJ, UK
\and Institut d'Astrophysique Spatiale (IAS), b\^atiment 121, Universit\'e Paris-Sud 11 and CNRS (UMR 8617), 91405 Orsay, France
\and Infrared Processing and Analysis Center, MS 100-22, California Institute of Technology, JPL, Pasadena, CA 91125, USA
\and School of Physics and Astronomy, The University of Manchester, Alan Turing Building, Oxford Road, Manchester M13 9PL, UK
\and Institut d'Astrophysique de Paris, UMR 7095, CNRS, UPMC Univ. Paris 06, 98bis boulevard Arago, F-75014 Paris, France
\and Mullard Space Science Laboratory, University College London, Holmbury St. Mary, Dorking, Surrey RH5 6NT, UK
\and Space Science \& Technology Department, Rutherford Appleton Laboratory, Chilton, Didcot, Oxfordshire OX11 0QX, UK
\and Institute for Space Imaging Science, University of Lethbridge, Lethbridge, Alberta, T1K 3M4, Canada
\and Astrophysics, Oxford University, Keble Road, Oxford OX1 3RH, UK
\and Warwick Systems Biology Centre, Coventry House, University of Warwick, Coventry CV4 7AL, UK
\and Centre for Astrophysics Research, University of Hertfordshire, College Lane, Hatfield, Hertfordshire AL10 9AB, UK}

   \date{Received Mar 30, 2010; accepted  May 11, 2010}

 
  \abstract
   {Emission at far-infrared wavelengths makes up a significant fraction of the total light detected from galaxies over the age of Universe. {Herschel} provides an  opportunity for studying galaxies at the peak wavelength of their emission.
   Our aim is to provide a benchmark for models of galaxy population evolution and to test pre-existing models of galaxies.
   With the {Herschel} Multi-tiered Extra-galactic survey, HerMES, we have observed a number of fields of different areas and sensitivity using the SPIRE instrument on {Herschel}.    
  We have determined the number counts of galaxies down to $\sim20$~mJy.  Our { constraints from directly counting galaxies} are consistent with, { though more precise than}, estimates from the BLAST fluctuation analysis. We have found a steep rise in the Euclidean normalised counts $<100$~mJy. We have directly { resolved $\sim15$\% of} the infrared extra-galactic background at the wavelength near where it peaks.}
   \keywords{galaxies: evolution, submillimeter: galaxies, surveys  }

   \maketitle

%

\section{Introduction} The statistical properties of galaxy populations are important probes for understanding the evolution of galaxies. The most basic statistic of galaxy populations is the number counts i.e. the number density of galaxies as a function of flux.  The first strong evidence for cosmological evolution came through studying number counts of radio galaxies (e.g. \citealt{longair66}).











The number counts at far-infrared and sub-mm wavelengths are well known to exhibit strong evolution, e.g. from {\em IRAS} 
\citep[and references therein]{Oliver92}, {\em ISO} \citep[and references therein]{Oliver2002,her04}
{\em Spitzer} \citep[and references therein]{Shupe2008, Frayer2009},
and ground-based sub-mm surveys \citep[and references therein]{Maloney2005,Coppin2006,Khan2007,Greve2008,Weiss2009,Scott2010}.

These results are underlined by the discovery of a significant extragalactic infrared background \citep{puget96,Fixsen1998,Lagache1999}. The background measures the flux weighted integral of the number counts over all redshifts plus any diffuse cosmological component.  This indicates that as much energy is received from galaxies after being reprocessed through dust as is received directly.  
It is only very recently, using BLAST, that { count models have} been probed { using fluctuation techniques \citep{Patanchon2009} or directly \citep{Bethermin2010}} at the wavelength where the background peaks.

{  Far-infrared and sub-mm counts and background measurements} have been modelled phenomenologically with strongly evolving populations (see Section \ref{sec:models}). Physical models (e.g. so-called semi-analytic models) struggle to explain these counts and solutions include altering the initial mass function (e.g. \citealt{Baugh2005})  { or exploiting AGN/Supernovae feedback (e.g. \citealt{Granato2004})}.

A primary goal of {\em Herschel} \citep{Pilbratt2010} is to explore the evolution of obscured galaxies. {\em Herschel} opens up a huge region of new parameter space of surveys in area, depth and wavelength.  

The {\em Herschel} multi-tiered extragalactic survey (HerMES\footnote{hermes.sussex.ac.uk}; \citealt{oliver2010}) is the largest project being undertaken by {\em Herschel} and consists of a  survey of many well-studied extra-galactic fields (totalling $\sim70\, {\rm deg}^2$) at various depths. This letter is the first number count analysis from the HerMES Science Demonstration Phase SPIRE  data. Even these preliminary results will be able to eliminate some existing models and provide a benchmark on which future models can be tested.

  
\section{SPIRE Data}\label{sec:data}
\subsection{Science Demonstration Phase Observations}
The observations described here were carried out on the {\em Herschel} Space Observatory \citep{Pilbratt2010} using the Spectral and Photometric Imaging Receiver (SPIRE). The SPIRE instrument, its in-orbit performance, and its scientific capabilities are described by \cite{Griffin2010}, and the SPIRE astronomical calibration methods and accuracy are outlined in \cite{Swinyard2010}. 
They were undertaken
as part of the HerMES programe during the Science Demonstration Phase between 12-Sep-2009 and 25-Oct-2009 under the proposal identification \verb+SDP_soliver_3+.   The fields and observations are summarised in Table \ref{tab:sdp_obs}. 

\begin{table*}
\caption{HerMES SPIRE SDP Observations. PACS observations will be discussed in \protect\cite{aussel2010}. Size is approximate extent of region with typical coverage. Roll angle is measured East of North. Repeats is total number of pairs of scans in both A and B directions. $t_{\rm AOR}$ is total time in execution of the observations. $\langle N_{\rm samp}\rangle$ is the mean number of bolometer samples per pixel in the same typical-coverage region of the 250 $\mu $m map ($6^{\prime\prime} \times 6^{\prime\prime}$pixels). $S_{50\%}^{250\mu{\rm m}}$ is the flux density at which 50\% of sources injected into the 250 $\mu $m map are faithfully recovered.}             
\label{tab:sdp_obs}      
\centering                          
\begin{tabular}{l c rrrllrrrrrrrrrr}        
\hline\hline                 
Name  &  Size &     RA & Dec & Roll & Mode & Scan  & Repeats & $t_{\rm AOR}$ & $\langle N_{\rm samp}\rangle$
& $S_{50\%}^{250\mu{\rm m}}$
& $S_{50\%}^{350\mu{\rm m}}$
& $S_{50\%}^{500\mu{\rm m}}$
\\ 
&  &    /$^\circ$ & /$^\circ$& /$^\circ$ & &  Rate & & /hr & & /mJy& /mJy& /mJy \\
\hline                        

A2218    &  $9^\prime\times9^\prime$ &  248.98   &    66.22      &  217 & Lrg. Map & 30$^{\prime\prime}$/s  &100 & 9.2 & 1622 & 13.8 & 16.0 & 15.1 \\

FLS      &  $155^\prime\times135^\prime$  &  258.97 &      59.39  &     185 & Parallel & 20$^{\prime\prime}$/s  & 1 &16.8 & 30 & 17.5 & 18.9 & 21.4\\
Lockman-North  &    $ 35^\prime\times35^\prime$   &    161.50 &      59.02    &   91 & Lrg. Map & 30$^{\prime\prime}$/s  & 7 & 3.9 & 117 & 13.7 & 16.5 & 16.0 \\
Lockman-SWIRE & $218^\prime \times 218^\prime$ & 162.00   &    58.11  &     92 & Lrg. Map & 60$^{\prime\prime}$/s  & 2 & 13.4 & 16 & 25.7 & 27.5 & 33.4 \\

GOODS-N     &   $30^\prime\times30^\prime $  &189.23   &    62.24 &      132 & Lrg. Map & 30$^{\prime\prime}$/s & 30 &13.5 & 501 & 12.0 & 13.7 & 12.8 \\
\hline                                   
\end{tabular}

\end{table*}

\subsection{SPIRE Catalogue Data Processing}
For this paper it is sufficient to note that the same source detection method is applied to the simulations as to the real data so we sketch the details only briefly. The single-band SPIRE catalogues have been extracted from the maps using a version of the {\sc sussex}tractor method \citep{Savage2007} as implemented in {\sc hipe} \citep{hipe2}. The processing of the SPIRE data is summarised here, with details of the approach given by \cite{Smith2010}. Calibrated timelines were created using HIPE development version 2.0.905, with a fix applied to the astrometry (included in more recent versions of the pipeline), with newer calibration files (beam-steering mirror calibration version 2, flux conversion version 2.3 and temperature drift correction version 2.3.2) and with a median and linear slope subtracted from each timeline. The default  {\sc hipe} na\"ive map-maker was then used to create maps, which were given a zero mean. 
The shallow fields { (Lockman-SWIRE and FLS)} were smoothed with a point-source optimised filter (see \citealt{Smith2010} for details). Peaks in the map were identified and the flux was estimated based on an assumed (Gaussian) profile for a point source, through a weighted sum of the map pixels close to the centre of the source.  { This filtering means we underestimate the flux of extended sources.}
{ 
The SPIRE Catalogue (SCAT) processing is assessed by injecting synthetic sources on a grid into the real maps. We then run the SCAT source extraction pipeline on these maps and claim success if the closest detection to the injected source is within a search radius of {\sc fwhm} { of the beam} and has a flux within a factor of two of the injected flux (see \citealt{Smith2010} for more details). The resulting 50\% completeness estimated in this way is tabulated in Table \ref{tab:sdp_obs} but is not used to assess the counts.}

\section{Method}\label{sec:method}
The {\em Herschel} beam is broad compared with the number density of sources, i.e. the maps are confused.  \citet{Nguyen2010} measure a variance in nominal map pixels due to confused sources finding $\sigma_{\rm conf}$ 5.8, 6.3 and 6.8 mJy/beam. This confusion means care has to be taken in the estimation of number counts.  
 Our technique follows the standard approach for sub-mm surveys, correcting for flux boosting and incompleteness.

We { determined the false detection rate by applying} the source extraction on maps obtained from the difference between two independent observations of the same field.  These maps are expected to have zero mean, { no sources}, but similar noise properties to mean maps. { We thus} estimate that the reliability for the samples in this paper is better than 97\% for all fields and bands.

A source we measure to have flux, $S_{\rm m}$, and noise, $\sigma_{\rm m}$, is more likely to be a dimmer source on top of a positive noise fluctuation than the converse; this is known as flux boosting.  We follow the Bayesian method of \cite{Crawford2009} for estimating fluxes of individual sources (``de-boosting''). We estimate the posterior probability distribution of the true flux of the source ($S_i$) that contributed the most flux to a given detection. Note that it is similar to the now-standard flux de-boosting method (e.g. \citealt{Coppin2006}) but with an additional  exponential suppression term at low intrinsic flux. 
We derive counts by randomly sampling the posterior distribution ten thousand times. The flux de-boosting procedure has some dependency on the choice of prior number count model and { so these samples are drawn from distributions produced for the full range of models discussed in Section \ref{sec:models}. These samples provide a direct estimate of the confidence region of our counts.}
For faint sources,  the posterior probability function rises beyond the sampled flux range ($S_{\rm m} / 5<S_i<5\,S_{\rm m}$) so that the deboosted flux is highly uncertain. In those cases we flag the deboosted flux as ``bad'' and the derived number counts at the flux level where deboosted fluxes are flagged ``bad' are unreliable.   { We exclude count bins in which the fraction of ``bad'' sources is $>20\%$}. 
We also estimate the uncertainty in this by looking at the variation in derived counts from the range of models.  Errors are included in the plots.
{ The flux de-boosting procedure assumes no clustering. Clustering will affect this and will be addressed in a later paper.}


{ We estimate the incompleteness in the whole process} by running full simulations.
We have constructed input maps various from various number count models (\citealt{pearson09, lagache04, Patanchon2009} and models from \citet{Xu2003} and \citet{Lacey2010}). These input maps are then processed by the SPIRE Photometer Simulator ({\sc sps}, \citealt{Sibthorpe2009}) for observational programmes exactly the same as { the real data}. The timeline output of the {\sc sps}, map-making and source extraction are then processed in the same way as the real data, { including the flux-deboosting.  The ratio of input to output counts gives us the completeness with the standard deviation between models providing an estimate of an error in that estimate}.






\section{Results}\label{sec:models}

\begin{figure}\centering
\includegraphics[width=3.4in]{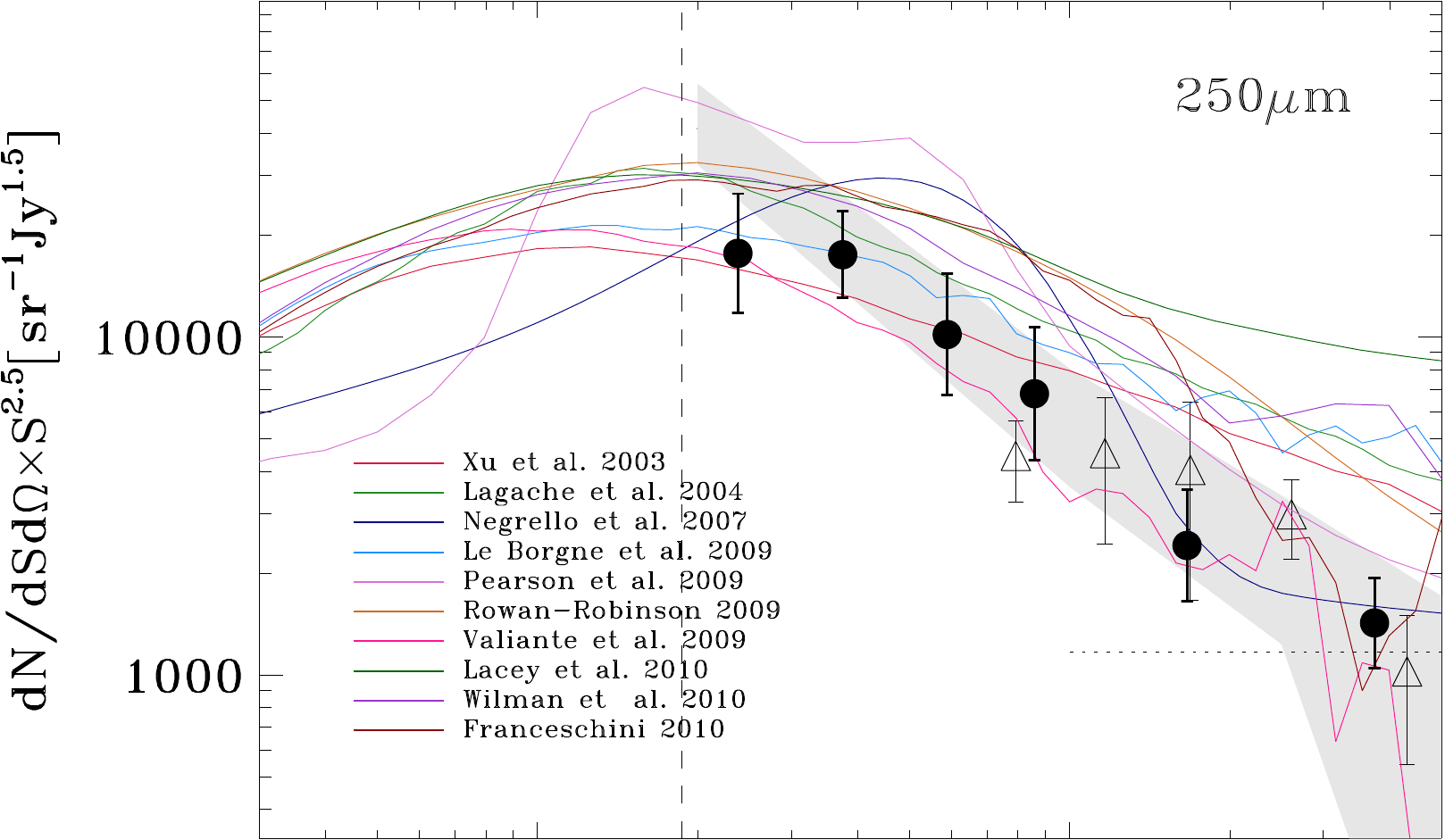}
\includegraphics[width=3.4in]{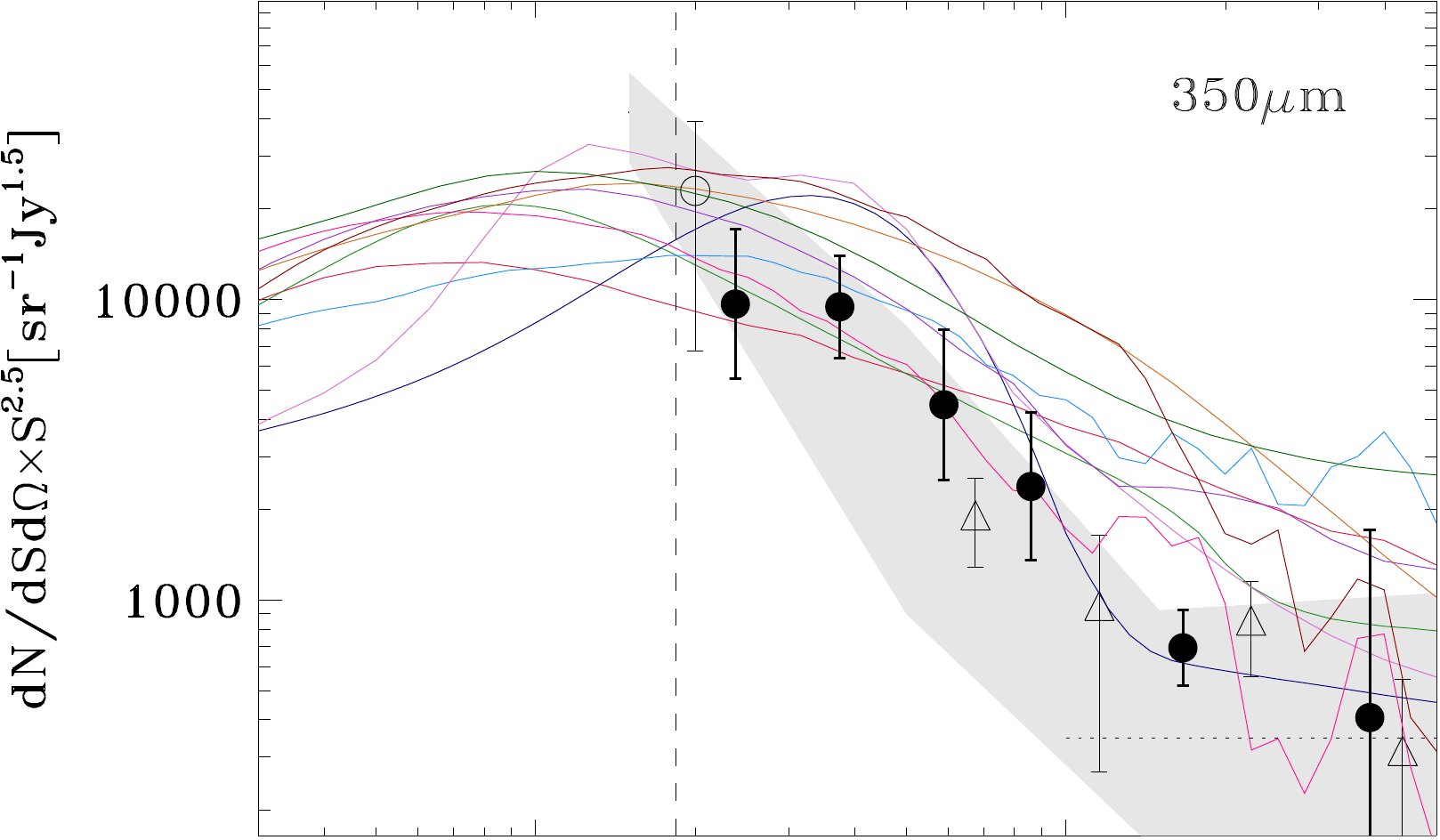}
\includegraphics[width=3.4in]{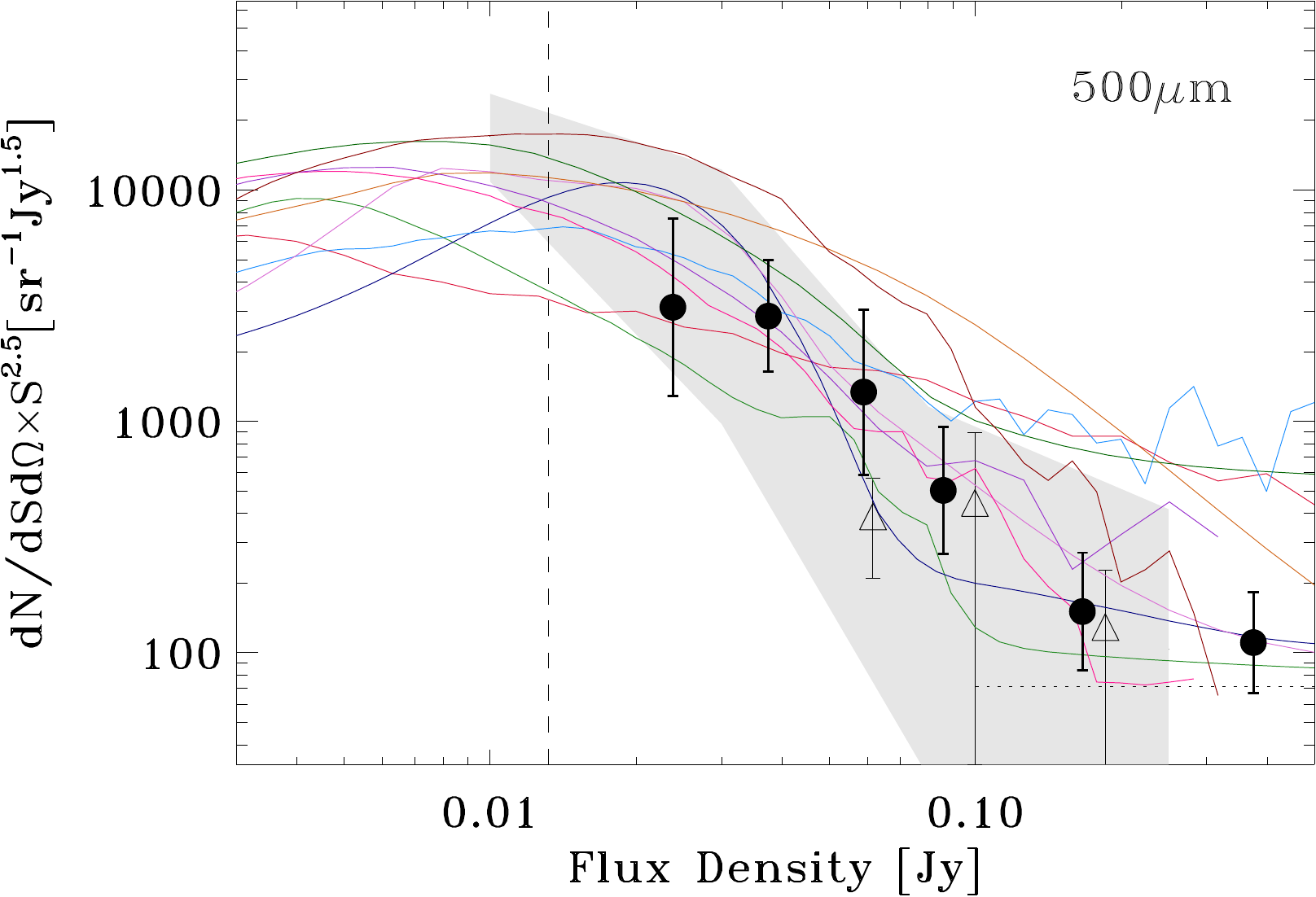}
\caption{Number counts obtained from HerMES source catalogues. 
Filled circles are the mean number counts averaged over the following fields. GOODS-N \& Lockman-North (faintest five bins only) and FLS \& Lockman-SWIRE (brightest six bins only) { with flux-deboosting, completeness corrections and field-field error bars.}  Model fit to fluctuations of BLAST maps (omitting upper-limits, \protect\citealt{Patanchon2009}) { --- shaded region; BLAST resolved counts \protect\citep{Bethermin2010} ---  open triangles; \protect\citet{Khan2007} data point --- open circle; asymptote from modelling of IRAS data \protect\citep{Serjeant2005} --- dotted line.  Models are discussed in the text.  Dashed line indicates the flux at which the integrated number density is (40 beams)$^{-1}$.}
}
\label{fig:realnumbercounts}
\end{figure}

\begin{figure}\centering
\includegraphics[width=3.4in]{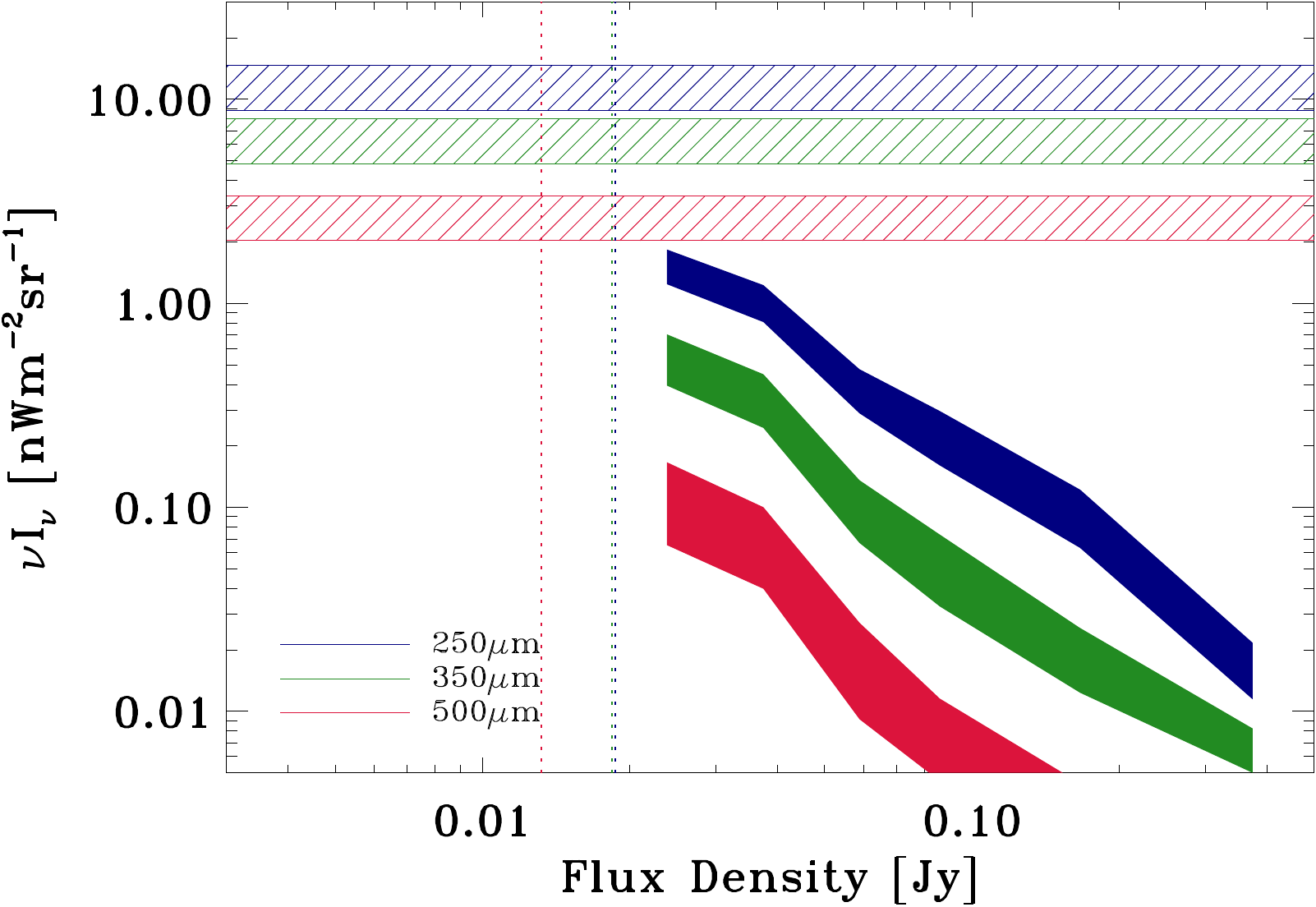}
\caption{The integrated background light at 250, 350, 500~$\mu$m from the HerMES counts determined in Figure~\protect\ref{fig:realnumbercounts}. Dotted lines { are the flux at which the integrated density is { (40 beams)$^{-1}$}}. The hatched regions are measurements of the COBE background \protect\citep{Lagache1999}}.
\label{fig:cib}
\end{figure}
{ The results are presented in Table \ref{tab:counts} and in Figure \ref{fig:realnumbercounts} with Euclidian normalization. 
There are several sources of uncertainties in the number counts: Poisson noise from the raw counts; ``sampling variance" due to additional fluctuations from real large-scale structure; additional Poisson noise from the sampling of the posterior flux distribution and systematic errors from the corrections and assumptions about priors and the effect of clustering on the de-boosting.  We measure the standard deviation of counts between fields which includes Poisson errors and some of the other systematic errors. The errors plotted are the field to field variations (or the Poisson errors if larger) with the errors from flux-boosting and completeness corrections added in quadrature.}

We see approximately flat counts for $S>100$~mJy and then a steep rise.  There is flattening to about 20~mJy.  We find very good agreement with the number counts estimated from a $P(D)$ fluctuation analysis of the BLAST maps \citep{Patanchon2009}.  

We have also estimated, but do not show, the integral counts. The flux density at which the integral source counts reach 1 source per 40 beams (with beams defined as $3.87\times10^{-5}, 7.28 \times10^{-5}, 1.48  \times 10^{-4}\, {\rm deg}^2$) is  { $18.7 \pm 1.2, \, 18.4 \pm 1.1$ and $13.2 \pm 1.0$~mJy} at 250, 350 and 500~$\mu$m respectively ({ N.B. these fluxes are slightly below our secure estimation of counts}) . Likewise the number density at 100~mJy is { $12.8\pm3.5,\, 3.7\pm 0.4$ and $0.8\pm0.1\, {\rm deg}^{-2}$}.  These last measurements alone will be sufficient to rule out many models.


Since the first {\em ISO} results, many empirical models have been developed to predict and interpret the numbers and luminosities of IR galaxies as a function of redshift. Empirical models are based on a similar philosophy. The spectral energy distributions of different galaxy populations are fixed and the mid-IR, far-IR and submm data are used to constrain the luminosity function evolution. Current limits come from the mid-IR, far-IR and submm number counts, redshift distributions, luminosity functions, and cosmic IR background. Models all agree on the general trends, with a very strong evolution of the bright-end ($> 10^{11} $\,L$_\odot$) of the luminosity function and they yield approximately the same comoving number density of infrared luminous galaxies as a function of redshift. We compare the number counts with eight models, one pre-{\em Spitzer} \citep{Xu2003}, two based on the {\em ISO}, SCUBA and {\em Spitzer} first results \citep{lagache04,negrello07} and 5 being more constrained by deep {\em Spitzer}, SCUBA, AzTEC, and recent BLAST observations \citep{LeBorgne2009,pearson09,rr09,Valiante2009,Franceschini2010}. The differences between the models are in several details, different assumptions leading sometimes to equally good fits to the current data. For example, \citet{Valiante2009} conclude that it is necessary to introduce both an evolution in the AGN contribution and an evolution in the luminosity-temperature relation, while \citet{Franceschini2010}  reproduce the current data with only 4 galaxy populations and only one template for each population. 
We also compare with two semi-analytic models those of \citet{Lacey2010} and \citet{Wilman2010}.

Comparison with SPIRE number counts shows that many models cannot fit the bright end ($>100$~mJy).  Exceptions are the models of \citet{negrello07}, \citet{Valiante2009}, \citet{Franceschini2010} and \citet{pearson09}. Of these only \citet{Valiante2009} can fit the rise from ($20<S<100$)~mJy.  { The \citet{Valiante2009} model has ``cooler'' spectral energy distributions at higher redshift. However, increasing the number of higher redshift galaxies would have a similar effect on the counts so it would be premature to assume the spectral energy distributions need revision.}

We have also calculated the contribution of the resolved sources to the background intensity as a function of flux (shown in Figure \ref{fig:cib}).  { At the (40 beams)$^{-1}$ depth we resolve $1.73\pm 0.33,\, 0.63\pm 0.18,\, 0.15\pm0.07\; {\rm nW}\,{\rm m}^2$ or 15, 10, 6\% of} the nominal measured values at 250, 350 and 500~$\mu$m \citep{Lagache1999}.

{ Future work will provide more detailed constraints at the fainter limits.  This will include a $P(D)$ analysis \citep{Glenn2010} and counts from catalogues extracted at known {\it Spitzer} source positions.}

\begin{table}
\caption{Table of HerMES SDP number counts at 250, 350 and 500$\mu$m. Bin limits and Euclidian weighted central fluxes are given in mJy.  Counts are in $/{\rm sr}^{-1}{\rm Jy}^{-1}$. Errors $1-4$ are fractional errors in percentages arising from: flux-deboosting, completeness corrections, Poisson statistics and field-field variations respectively.}\label{tab:counts}
\begin{tabular}{ccccccccc}
\hline
\hline
   & Bin 1 & Bin 2 & Bin 3 & Bin 4 & Bin 5 & Bin 6 \\
\hline
                      $S_{\rm min}$&    20 &    29 &    51 &    69 &   111 &   289 \\
                       $S_{\rm max}$&    29 &    51 &    69 &   111 &   289 &   511 \\
                       $S_{\rm euc}$&  23.8 &  37.5 &  58.9 &  85.9 & 166.2 & 374.1 \\
\hline
\multicolumn{7}{c}{250$\mu$m}\\

      $\frac{dN}{dS}$& 2.0$\times 10^8$ & 6.4$\times 10^7$ & 1.2$\times 10^7$ & 3.1$\times 10^6$ & 2.1$\times 10^5$ & 1.7$\times 10^4$ \\
                              err$_1$&    38 &    22 &    28 &    35 &    23 &     6 \\
                              err$_2$&    10 &     6 &     7 &     6 &     4 &    19 \\
                              err$_3$&     4 &     3 &     7 &     4 &     7 &    23 \\
                              err$_4$&    11 &    19 &    30 &    28 &    30 &     8 \\
\hline
\multicolumn{7}{c}{350$\mu$m}\\
      $\frac{dN}{dS}$ & 1.1$\times 10^8$ & 3.5$\times 10^7$ & 5.3$\times 10^6$ & 1.1$\times 10^6$ & 6.2$\times 10^4$ & 4.7$\times 10^3$ \\
                              err$_1$&    49 &    34 &    44 &    56 &    25 &   129 \\
                              err$_2$&    18 &    14 &    17 &     6 &     5 &    12 \\
                              err$_3$&     7 &     4 &    10 &     7 &    15 &    43 \\
                              err$_4$&    23 &    13 &    33 &     8 &    12 &    64 \\
\hline
\multicolumn{7}{c}{500$\mu$m}\\
      $\frac{dN}{dS} $ & 3.6$\times 10^7$ & 1.1$\times 10^7$ & 1.6$\times 10^6$ & 2.3$\times 10^5$ & 1.3$\times 10^4$ & 1.3$\times 10^3$ \\
                              err$_1$&    83 &    50 &    62 &    56 &    45 &     0 \\
                              err$_2$&    31 &    18 &    25 &    15 &    18 &     7 \\
                              err$_3$&    10 &     6 &    18 &    14 &    33 &    50 \\
                              err$_4$&     5 &    17 &    48 &    27 &    20 &     0 \\
\hline
\end{tabular}
\end{table}

\section{Conclusions}\label{sec:conclusions}

{  We present the first SPIRE number count analysis of resolved sources, conservatively within the limit of {\em Herschel} confusion.  We have measured counts which resolve around 15\% of the infrared background at 250~$\mu$m.  We see a very steep rise in the counts from 100 to 20~mJy in 250, 350 and 500~$\mu$m. Few models have quite such a steep rise. Many models fail at the bright counts $>100$~mJy.  This may suggest that models need a wider variety or evolution of the spectral energy distributions or changes in the redshift distributions.  Future work is required to accurately constrain the fainter ends $<20$~mJy where confusion is a serious challenge.  }

\begin{acknowledgements}

      Oliver, Wang and Smith were supported by UK's Science and Technology Facilities Council grant ST/F002858/1.
SPIRE has been developed by a consortium of institutes led by
Cardiff Univ. (UK) and including Univ. Lethbridge (Canada);
NAOC (China); CEA, LAM (France); IFSI, Univ. Padua (Italy);
IAC (Spain); Stockholm Observatory (Sweden); Imperial College
London, RAL, UCL-MSSL, UKATC, Univ. Sussex (UK); Caltech, JPL,
NHSC, Univ. Colorado (USA). This development has been supported
by national funding agencies: CSA (Canada); NAOC (China); CEA,
CNES, CNRS (France); ASI (Italy); MCINN (Spain); SNSB (Sweden);
STFC (UK); and NASA (USA).
HIPE is a joint development (are joint developments) by the Herschel Science Ground
Segment Consortium, consisting of ESA, the NASA Herschel Science Center, and the HIFI, PACS and
SPIRE consortia."
      The data presented in this paper will be released through the {\em Herschel} Database in Marseille HeDaM ({hedam.oamp.fr/HerMES}).
      
      \end{acknowledgements}
\bibliography{counts}
\end{document}